\documentclass[11pt,letterpaper]{scrartcl}

\usepackage{scicite}
\usepackage[varg]{txfonts}
\usepackage{graphicx}

\newenvironment{sciabstract}{%
\begin{quote} \bf}
{\end{quote}}

\newcounter{lastnote}
\newenvironment{scilastnote}{%
\setcounter{lastnote}{\value{enumiv}}%
\addtocounter{lastnote}{+1}%
\begin{list}%
{\arabic{lastnote}.}
{\setlength{\leftmargin}{.28in}}
{\setlength{\labelsep}{.5em}}}
{\end{list}}

\title{A Common Explosion Mechanism for Type Ia Supernovae}

\author
{Paolo~A.~Mazzali$^{1,2,3,4\ast}$,
Friedrich K.~R\"opke$^{1,5}$,\\
Stefano Benetti$^{6}$,
and Wolfgang Hillebrandt$^{1}$
\\
\\
\normalsize{$^{1}$Max-Planck Institut f\"ur Astrophysik,}\\
\normalsize{Karl-Schwarzschildstr.1, 85748 Garching, Germany}\\
\normalsize{$^{2}$Department of Astronomy, School of Science,}\\ 
\normalsize{University of Tokyo, Bunkyo-ku, Tokyo 113-0033, Japan}\\
\normalsize{$^{3}$Research Center for the Early Universe, School of Science,}\\
\normalsize{University of Tokyo, Bunkyo-ku, Tokyo 113-0033, Japan}\\
\normalsize{$^{4}$Istituto Nazionale di Astrofisica-OATs, Via Tiepolo 11,
I-34131 Trieste, Italy}\\ 
\normalsize{$^{5}$Department of Astronomy and Astrophysics, University
of California Santa Cruz,}\\
\normalsize{1156 High Street, Santa Cruz, CA 95064, U.S.A.}\\ 
\normalsize{$^{6}$Istituto Nazionale di Astrofisica-OAPd,}\\ 
\normalsize{vicolo dell'Osservatorio, 2, I-35122 Padova, Italy}\\
\\
\normalsize{$^\ast$To whom correspondence should be addressed; E-mail: mazzali@ts.astro.it}\\
}

\date{}

\def\gsim{\mathrel{\rlap{\lower 4pt \hbox{\hskip 1pt $\sim$}}\raise 1pt \hbox {$>$}}} 
\def\lsim{\mathrel{\rlap{\lower 4pt \hbox{\hskip 1pt $\sim$}}\raise 1pt \hbox {$<$}}}

\def\aj{AJ}%
\def\araa{ARA\&A}%
\def\apj{ApJ}%
\def\apjl{ApJ}%
\def\aap{A\&A}%
\def\mnras{MNRAS}%
\def\nat{Nature}%
\begin{document} 




\maketitle


\begin{sciabstract}
Type Ia supernovae, the thermonuclear explosions of white dwarf stars
composed of carbon and oxygen, were instrumental as distance
indicators in establishing the acceleration of the universe's
expansion. However, the physics of the explosion are debated.  Here we
report a systematic spectral analysis of a large sample of well
observed type Ia supernovae. Mapping the velocity distribution of the
main products of nuclear burning, we constrain theoretical
scenarios. We find that all supernovae have low-velocity cores of
stable iron-group elements. Outside this core, nickel-56 dominates the
supernova ejecta. The outer extent of the iron-group material depends
on the amount of nickel-56 and coincides with the inner extent of
silicon, the principal product of incomplete burning. The outer extent
of the bulk of silicon is similar in all SNe, having an expansion
velocity of $\sim 11000\, \mathrm{km}\,\mathrm{s}^{-1}$ and corresponding
to a mass of slightly over one solar mass. This indicates that all the
supernovae considered here burned similar masses, and suggests that
their progenitors had the same mass. Synthetic light curve parameters
and three-dimensional explosion simulations support this
interpretation. A single explosion scenario, possibly a delayed
detonation, may thus explain most type Ia supernovae.
\end{sciabstract}

When a white dwarf (WD) composed of carbon and oxygen accreting mass
from a companion star in a binary system approaches the Chandrasekhar
mass [$M_{\mathrm{Ch}} \approx 1.38$ solar masses ($M_{\odot}$)], high
temperature causes the ignition of explosive nuclear burning reactions
that process stellar material and produce energy. The star explodes
leaving no remnant, producing a Type Ia supernova (SNIa)
\cite{nom84}. At high stellar material densities, burning yields
nuclear statistical equilibrium (NSE) isotopes, in particular
radioactive $^{56}$Ni which decays to $^{56}$Co and $^{56}$Fe making
the SN bright \cite{kuch94}. At lower densities intermediate mass
elements (IME) are synthesised. Both groups of elements are observed
in the optical spectra of SNeIa \cite{bra85}. An empirical relation
between an observed quantity, the B-magnitude decline over the first
15 days after maximum luminosity [$\Delta m_{15} (B)$] and a physical
quantity, the SN maximum luminosity \cite{phil93}, can be used to
determine the distance to a SNIa. This method was applied to very
distant SNeIa leading to the discovery of the accelerating Universe
\cite{riess98,perl99}. How the explosion actually proceeds is however
debated \cite{reinecke02,gamezo04,plewa04}, as is the nature of the
progenitor system: Accretion may occur either from a more massive
companion (e.g. a giant) or via the merging of two carbon-oxygen WDs
\cite{hillnie00}. This casts a shadow on the reliability of SNeIa as
distance indicators, as intrinsically very different explosions may
result in the observed correlation.

We derive the distribution of the principal elements in 23 nearby
SNeIa (distances $<$$40 \, \mathrm{Mpc}$) with good spectral coverage
extending from before maximum to the late nebular phase, about one
year later. The sample (table S1) covers a wide range of light-curve
decline rates and includes peculiar objects like SN2000cx, which
violates the luminosity-decline-rate relation.

Because of the hydrodynamic properties of the explosion, the expansion
velocity of the ejecta is proportional to radius and serves as a
radial coordinate. As the SN expands, deeper layers are exposed.  The
outer layers, visible in the first few weeks after the explosion, are
dominated by IME. Because silicon is the most abundant IME, we measured its
characteristic expansion velocity from the blueshift of the absorption
core of the strong SiII $6355\,\mbox{\AA}$ line in all spectra where it was
visible.  This velocity decreases with time. Fitting the postmaximum
velocity evolution and extrapolating it to the earliest times, when
the outermost parts of the ejecta are visible, we derived the outer
extension of the bulk of Si. This represents a lower limit of the
outer extent of burning. The Si velocity
$v(\mathrm{Si})$ is similar in all SNeIa, regardless of their
luminosity: $v(\mathrm{Si}) = 11900 \pm 1300 \, \mathrm{km} \,
\mathrm{s}^{-1}$.

A few SNe, defined as High Velocity Gradient (HVG) SNe \cite{ben05}, are
responsible for most of the dispersion.  They have a rapidly
decreasing $v$(Si) before maximum and very high-velocity CaII lines,
possibly the result of high-velocity blobs that carry little mass and
kinetic energy but cause High-Velocity absorption Features (HVF,
\cite{maz05}) that can abnormally broaden the SiII line profile
\cite{maz05,tan06}. Although excluding them from the sample decreases the
dispersion in $v$(Si) significantly [$v(\mathrm{Si}) = 11300 \pm 650
\, \mathrm{km} \, \mathrm{s}^{-1}$], we include all these SNe in our
discussion.

The maximum Si velocity thus measured is a conservative
estimate. The deep absorption core is produced in layers of high
silicon abundance.  Silicon is present at higher velocities, indicated
by the wavelength of bluest absorption in the earliest spectrum of
each SN.  However, measuring the bluest absorption velocity yields a
large scatter because the earliest spectra have different epochs and
may be affected by HVFs \cite{maz05}. Our method reliably determines the outer
location of the bulk of IMEs. 

\begin{figure}[t]
\centerline{\includegraphics[width=0.65\linewidth]{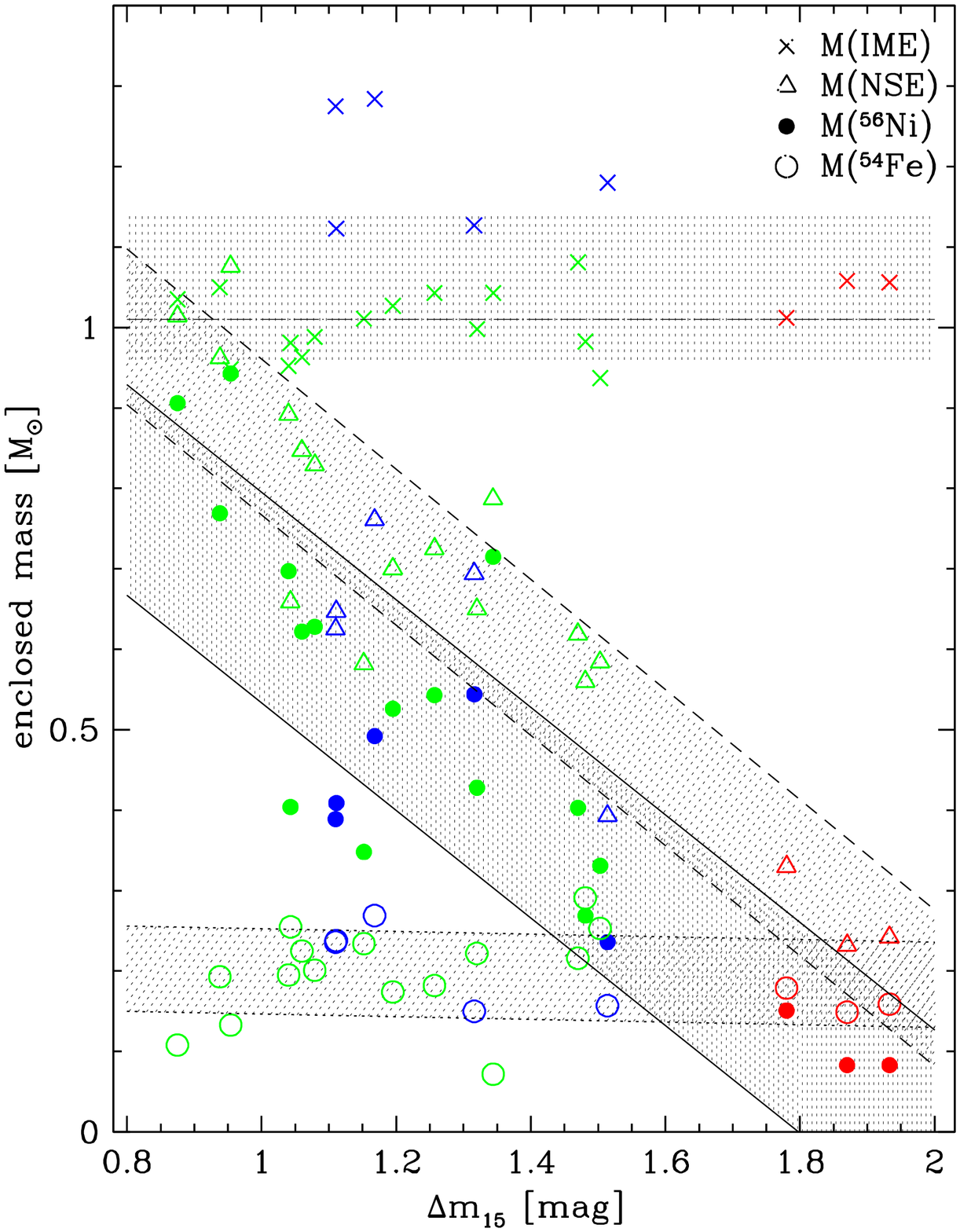}}
\caption{\footnotesize The Zorro Diagram: Distribution of the
  principal isotopic groups in SNeIa.  The enclosed mass (linked to
  velocity via the W7 explosion model) of different burning products
  is shown versus decline-rate parameter $\Delta m_{15} (B)$ (a proxy
  for SN luminosity).  Individual SNe are coloured according to their
  velocity evolution \cite{ben05}: HVG, blue; low
  velocity gradient (LVG), green; and faint, red.  Open circles indicate
  the mass of stable $^{54}$Fe+$^{58}$Ni for each SN; solid circles
  that of $^{56}$Ni, and open triangles the sum of these (total NSE
  mass). Crosses show the sum of NSE and IME mass, indicating the
  total mass burned. The IME mass is the difference between crosses
  and triangles.  $^{54}$Fe and $^{58}$Ni are found in roughly
  constant amounts in the deepest parts of all SNe, irrespective of
  luminosity: $M (\mathrm{stable NSE}) = 0.20 \pm 0.05 \, M_{\odot}$
  (lower horizontal shaded area).  The $^{56}$Ni mass determines the
  SN luminosity. It correlates with $\Delta m_{15} (B)$:
  $M(^{56}\mathrm{Ni}) = 1.34 - 0.67 \Delta m_{15} (B) [M_{\odot}]$,
  with rms dispersion $0.13 \, M_{\odot}$ (lower diagonal shaded
  area).  The total NSE mass correlates with $\Delta m_{15} (B)$
  better than M($^{56}$Ni): $M(\mathrm{NSE}) = 1.55 - 0.69 \Delta
  m_{15} (B) [M_{\odot}]$, rms dispersion $0.09\, M_{\odot}$ (upper
  diagonal shaded area).  IME lie mostly outside the iron-group zone.
  The outer Si velocity is similar for all SNe except HVG SNe.  The
  mass enclosed by IMEs represents the total burned mass. When all SNe
  are included the average value is $\sim$$1.05 \pm 0.09 \, M_{\odot}$
  (upper horizontal shaded area).  Excluding HVG SNe the value is
  $\sim$$1.01 \pm 0.05 \, M_{\odot}$ (horizontal line).  Both values
  are independent of $\Delta m_{15} (B)$.  For a version of this plot
  with SN names and a bar diagram, see figs. S2 and S3.
\label{fig:Zorro}}
\end{figure}

The inner extent of Si, determined from the asymptotic velocity
of the SiII 6355$\, \mbox{\AA}$ line in post-maximum spectra
[fig.~1 in \cite{ben05}], is a steep function of $\Delta
m_{15} (B)$. The brightest (slowest declining) SNe have the thinnest
Si zones.

The inner ejecta, dominated by NSE elements, are best observed $\sim$1
year after the explosion, when dilution caused by expansion makes the
SN behave like a nebula, exposing the deepest layers.  Collisions with
the fast particles produced by the decay
$^{56}$Ni$\,\to\,$$^{56}$Co$\,\to\,$$^{56}$Fe heat the gas, which
cools emitting radiation mostly in forbidden lines.

We modeled the nebular spectra using a code that computes line
emission balancing heating and cooling in non-local thermodynamic
equilibrium (NLTE) \cite{maz01a}, including density and abundance
stratification. We adopted the density-velocity distribution of the
standard, one-dimensional, $M_\mathrm{Ch}$ explosion model W7
\cite{nom84}, (fig. S1). In the nebular phase the gas is transparent,
and line emissivity depends on the mass of the emitting
ion. Accurately estimating this mass requires determining the
ionization state of the gas. Forbidden lines of FeII and FeIII
dominate SNIa nebular spectra, reflecting the high abundance of NSE
material. Fe is mostly the product of $^{56}$Ni decay, which provides
heating. The stable, neutron-rich isotopes $^{54}$Fe and $^{58}$Ni do
not contribute to heating, but do contribute to cooling, because they
also emit forbidden lines. Their presence affects the ionization
balance. Both are mostly produced deep in the WD, at the highest
densities. The $^{54}$Fe nebular lines have wavelengths
indistinguishable from those of $^{56}$Fe.  

We determined the mass and
distribution in velocity of the Fe isotopes (and thus of $^{56}$Ni)
simultaneously fitting the ratio of the two strongest Fe
emissions(fig.~S1). One, near 4700$\, \mbox{\AA}$ includes both FeII and
FeIII lines, while the other, near 5200$\, \mbox{\AA}$, is only due to
FeII. A low upper limit to the mass of $^{58}$Ni is set by the
absence of strong emission lines (in particular at 7380$\, \mbox{\AA}$).

In the deepest layers (Fig.~\ref{fig:Zorro}), all SNe contain
$\sim$0.1 to $0.3 \, M_{\odot}$ of stable NSE isotopes, with a large
scatter and no dependence on $\Delta m_{15} (B)$ [see also
\cite{woo06}]. 

As expected, the $^{56}$Ni mass correlates inversely with $\Delta
m_{15} (B)$, ranging from $0.9 \, M_{\odot}$ for the slowest declining
(most luminous) SNe to $0.1\, M_{\odot}$ for the fastest declining
(dimmest) ones. The root mean square (rms) dispersion is $0.13
\,M_{\odot}$, but SNe with intermediate decline rates [$\Delta m_{15}
(B) \sim 1.05$ to $1.5~\mathrm{magnitudes}$] show variations of almost a factor of 2
for the same value of $\Delta m_{15} (B)$.  These SNe could cause
scatter about the mean luminosity-decline-rate relation \cite{maz06}.

Once the contributions of $^{56}$Ni, $^{54}$Fe, and
$^{58}$Ni are added together to evaluate the total NSE mass, the
dispersion decreases to $0.09\, M_{\odot}$. If SNe with different
amounts of $^{56}$Ni, and thus presumably different temperatures, but
similar NSE content have similar $\Delta m_{15} (B)$, it is likely
that abundances \cite{maz01} rather than temperature \cite{kasen06}
primarily determine the opacity and light-curve shape.

The outer velocity of the NSE region, determined from the width of the
Fe lines, correlates with SN luminosity \cite{maz98}. It coincides with the
innermost silicon velocity, marking the transition from complete to
incomplete burning: the difference between these two velocities,
$\Delta(v) = 650 \pm 900 \, \mathrm{km} \, \mathrm{s}^{-1}$, is
consistent with zero. Remarkably, this results from two different
methods applied to data obtained almost 1 year apart. 

Thus, while the mass of $^{56}$Ni, and consequently the SN luminosity, can
differ significantly, other characteristics of SNeIa are remarkably
homogeneous.  In particular, the narrow dispersion of the outer Si
velocity indicates a similar extent of thermonuclear burning in all
SNeIa: SNe that produce less $^{56}$Ni synthesize more IMEs. 

The simplest interpretation of these seemingly antithetic results is
that thermonuclear burning consumes similar masses in all SNeIa.  We
explore whether this scenario is consistent with the $M_{\mathrm{Ch}}$
model \cite{hillnie00}.  Applying the density-velocity structure of
model W7 \cite{nom84}, we transform $v$(Si) to mass
(Fig.~\ref{fig:Zorro}). We find that the outer shell of silicon
encloses a mass of at least $\sim 1.05 \pm 0.09 \, M_{\odot}$ ($\sim
1.01 \pm 0.05 \, M_{\odot}$, excluding HVG SNe), independently of
$\Delta m_{15} (B)$. This is a lower limit to the burned mass.

\begin{figure}[t]
\centerline{\includegraphics[width=0.75\linewidth]{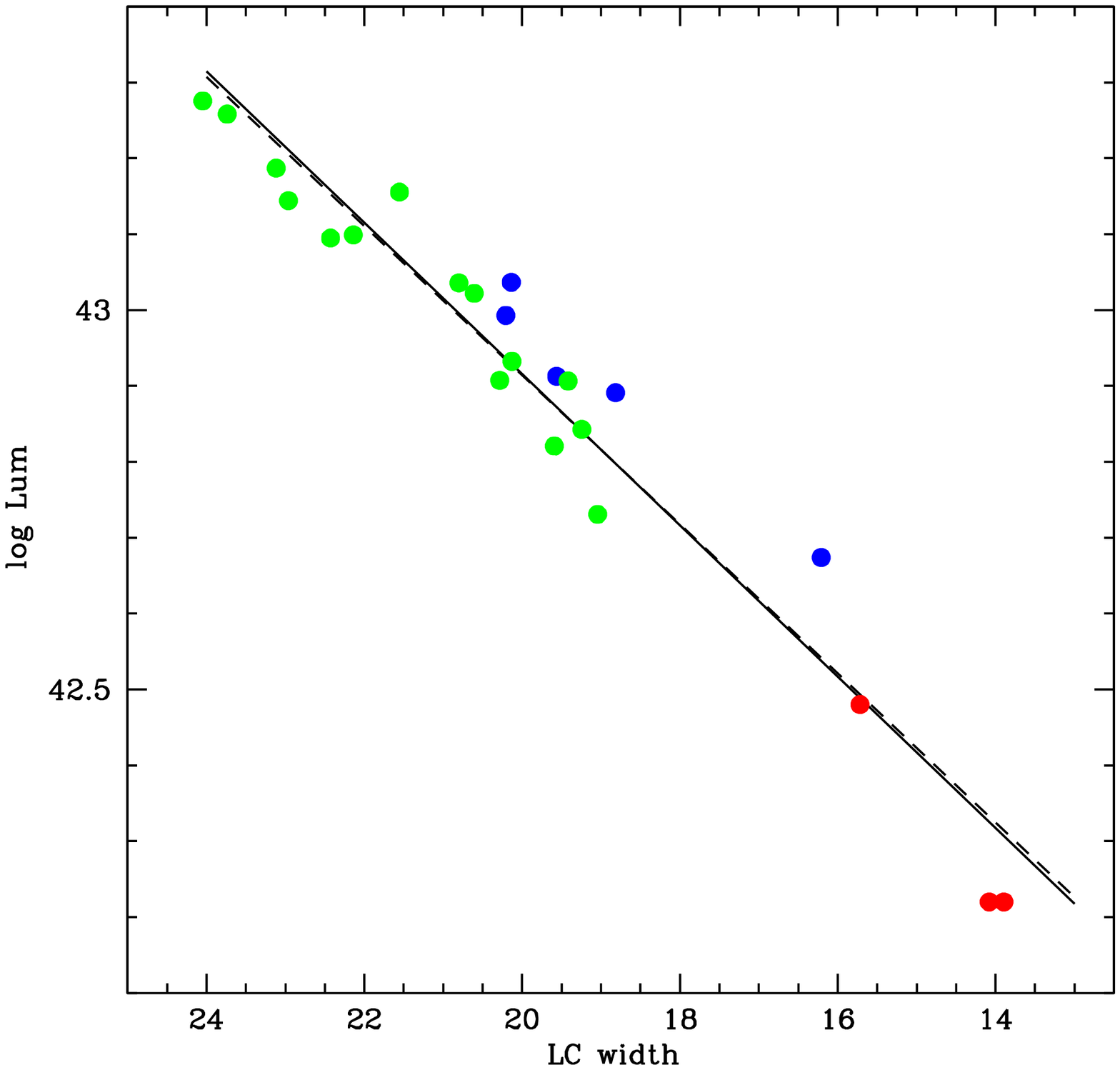}}
\caption{\footnotesize Observed and synthetic luminosity-decline-rate
relations for the SNe in our sample. Colors indicate velocity
evolution as in Fig.~\ref{fig:Zorro}.  The peak luminosity $L$ was computed from
M($^{56}$Ni) as $L = 2 \times 10^{43} \, M(^{56}\mathrm{Ni})$
\cite{stritz}.  Observed bolometric light-curve (LC) widths were
obtained from observed rise and decline times \cite{contardo00} as $T
= (t_{-1/2}+ t_{+1/2})$.  When $t_{-1/2}$ is missing, it was estimated
from the relation between $t_{-1/2}$ and $\Delta m_{15} (B)$ derived
from the other SNe.  The dashed line is a linear regression between
luminosity and observed LC width for the seven SNe common to our sample
and that of \cite{contardo00}.  Synthetic bolometric LC widths were
obtained assuming that: (i) The LC width $\tau$ depends on ejected mass
$M_{\mathrm{ej}}$, kinetic energy $E_{\mathrm{k}}$, and opacity
$\kappa$ as: $\tau \propto \kappa^{1/2} E_{\mathrm{k}}^{-1/4}
M_{\mathrm{ej}}^{3/4}$ \cite{arn82}. (ii) For all SNe, $M_{\mathrm{ej}}
= M_{\mathrm{Ch}}$. The W7 \cite{nom84} density-velocity distribution
was used. (iii) The explosion kinetic energy depends on the burning
product: $ E_{\mathrm{k}} = [1.56 M(^{56}\mathrm{Ni}) + 1.74
M(\mathrm{stable NSE}) + 1.24 M(\mathrm{IME}) - 0.46] \times
10^{51}\,\mathrm{erg}$ \cite{maz01}. (iv) Opacity is mostly due to line
absorption \cite{paul97}. Accordingly, NSE elements contribute much
more than IMEs because their atomic level structure is more complex. The
opacity was therefore parametrised according to the abundances of
different species: $\kappa \propto M(\mathrm{NSE}) + 0.1\,
M(\mathrm{IME})$ \cite{maz06}.  To compare the parametrized LC widths
$\tau$ to observed values T, a scale factor $x=T/\tau$ was computed
for each of the seven SNe common to our sample and that of
\cite{contardo00}. The average factor, $x=24.447$, was used to scale
all SNe. Dots show the individual SNe. The continuous line is the
linear regression between luminosity and synthetic LC width for our 23
SNeIa sample.  For a version of this plot with SN names, see fig.~S4.
\label{fig:LCprops}}
\end{figure}

The light-curve width $\tau$ is related to $\Delta m_{15} (B)$
\cite{gold01} and depends on the ejected mass $M_{\mathrm{ej}}$, the
kinetic energy $E_{\mathrm{k}}$, and the opacity $\kappa$ as $\tau
\propto \kappa^{1/2} E_{\mathrm{k}}^{-1/4} M_{\mathrm{ej}}^{3/4}$
\cite{arn82}. We test our assumption computing parametrized
light-curve widths and comparing them to observed values. The
resulting luminosity-decline-rate relation (Fig.~\ref{fig:LCprops}) is
very tight, and practically identical to the observed one.  This
result supprots our opacity parametrization, corroborating our
hypothesis that the mass burned is similar in all SNeIa. Given its
weak dependence on $E_{\mathrm{k}}$, the light-curve width is not much
affected if more of the outer part of the WD is burned to IME, as
these do not contribute much to the opacity.

If SNeIa burn a similar mass, the progenitor mass is also likely to
be the same, namely $M_{\mathrm{Ch}}$. Because the outcome of the
burning depends essentially on fuel density, a variation in iron-group
elements production in $M_{\mathrm{Ch}}$ WDs requires different WD
expansion histories.  This in turn depends on the details of the
burning. Once the WD reaches $M_{\mathrm{Ch}}$, a thermonuclear flame
is ignited near the center.  The flame must start as a subsonic
deflagration \cite{nom84}, mediated by microphysical transport and
accelerated by turbulence. As it propagates outwards, it could undergo
a deflagration-to-detonation transition (DDT) and continue as a
shock-driven supersonic detonation wave, in a so-called delayed
detonation \cite{khokhlov91}. This constitutes the most extreme
explosion scenario admissible, exploring the limits of
$M_{\mathrm{Ch}}$ explosions.

\begin{table}[t]
\caption{Results and DDT parameters of parametrized three-dimensional delayed detonation
  models. \label{tab:res}}
\centerline{\begin{tabular}{llllll}
\hline
model  & $E_{\mathrm{kin}}^{\mathrm{asympt}} \, (\mathrm{10}^{51} erg)$ & $M(\mathrm{NSE}) \,
(M_\odot$) & $M(\mathrm{IME}) \, (M_\odot)$ & $t_{\mathrm{DDT}}$ (s) &
$\rho_{\mathrm{DDT}}$ ($10^7 \, \mathrm{g} \, \mathrm{cm}^{-3}$)\\
\hline
\emph{D800} & 1.004 & 0.638 & 0.547 & 0.675 & 2.40\\
\emph{D20}  & 1.237 & 0.833 & 0.435 & 0.724 & 1.92\\
\emph{D5}   & 1.524 & 1.141 & 0.220 & 0.731 & 1.33\\
\hline
\end{tabular}}
\end{table}

We modeled the explosion using a three-dimensional level-set approach
\cite{golombek05}. 
The ignition of the deflagration flame was treated as a stochastic
process generating a number of ignition spots placed randomly and
isotropically within $180\, \mathrm{km}$ of the WD centre \cite{woos}. 

What would cause the DDT is unclear. We assumed that it occurs as
turbulence penetrates the internal flame structure: the onset of the
so-called distributed burning regime \cite{niewoo97}. This happens at low fuel
densities, after some WD pre-expansion in the deflagration phase. The
detonation is triggered artificially where the chosen DDT criterion is
first satisfied, typically near the outer edges of the deflagration
structure (Table~\ref{tab:res} shows DDT parameters). 
Three simulations, with 800, 20, and 5 ignition spots, termed D800,
D20, and D5, respectively, were performed on a moving cellular
Cartesian grid \cite{roep06} comprising the full star. Model D800, with its
dense distribution of ignition points, exhausts the carbon-oxygen fuel
at the WD centre almost completely in the deflagration phase. The
energy release quickly expands the star, and the subsequent detonation
mainly transforms low density outer material to IME. In contrast, the
few ignition spots of model D5 consume little material during the
deflagration, leaving more fuel at high densities which is converted
mostly to NSE isotopes in the vigorous detonation phase. Model D20 provides an
intermediate case. 

The ejecta compositions of the model explosions agree grossly with the
results derived from the spectra. The NSE mass produced ranges from
0.638 to 1.141$\, M_{\odot}$.  The weakest explosions result from
optimal burning in the deflagration phase \cite{roep06}, and
subluminous SNeIa are not reached in our parametrization. However,
the conditions for DDT need further investigation. Eventually these
events may be explained within a single framework. Some extremely
luminous SNeIa may come from very rapidly rotating WDs whose mass
exceeds \cite{how06}, but these are rare.

The distribution derived from observations of burning products inside
SNeIa could result from the variation of a single initial parameter,
the flame ignition configuration, in M$_{\mathrm{Ch}}$ delayed detonations.  The
luminosity-decline-rate relation can be reproduced using this
distribution and a simple opacity parametrization. Our results support
the M$_{\mathrm{Ch}}$ scenario for most SNeIa, adding confidence to their use as
distance indicators.

\begin{scilastnote}
\item We thank Elena Pian and Daniel Sauer for help with data analysis.\\
This work was partly supported by the European Union's Human Potential
Programme under contract 
HPRN-CT-2002-00303, ``The Physics of Type Ia Supernovae''. 
\end{scilastnote}


\end{document}